\documentclass[11pt, fullpage]{article}
\usepackage{fullpage}
\usepackage{setspace}
\usepackage{authblk}
\usepackage[colorlinks,citecolor=blue]{hyperref}

\usepackage{cite}
\usepackage{amsmath,amssymb,amsfonts}
\usepackage{algorithmic}
\usepackage{graphicx}
\usepackage{textcomp}
\usepackage[caption=false]{subfig}
\usepackage[flushleft]{threeparttable}
\def\BibTeX{{\rm B\kern-.05em{\sc i\kern-.025em b}\kern-.08em
		T\kern-.1667em\lower.7ex\hbox{E}\kern-.125emX}}    
\allowdisplaybreaks
\usepackage{multirow}
\usepackage{xcolor}

\makeatletter
\usepackage{algorithm}
\newcommand\fs@norules{\def\@fs@cfont{\bfseries}\let\@fs@capt\floatc@ruled
	\def\@fs@pre{}%
	\def\@fs@post{}%
	\def\@fs@mid{\kern3pt}%
	\let\@fs@iftopcapt\iftrue}
\makeatother
\floatstyle{norules}
\restylefloat{algorithm}

\begin{document}

\title{Efficient Discontinuous Galerkin Scheme for Analyzing Nanostructured Photoconductive Devices}

\author{Liang Chen, Kostyantyn Sirenko, Ping Li, and Hakan Bagci}

\maketitle

	\begin{abstract}
		Incorporation of plasmonic nanostructures in the design of photoconductive devices (PCDs) has significantly improved their optical-to-terahertz conversion efficiency. However, this improvement comes at the cost of increased complexity for the design and simulation of these devices. Indeed, accurate and efficient modeling of multiphysics processes and intricate device geometries of nanostructured PCDs is challenging due to the high computational cost resulting from multiple characteristic scales in time and space. In this work, a discontinuous Galerkin (DG)-based unit-cell scheme for efficient simulation of PCDs with periodic nanostructures is proposed. The scheme considers two physical stages of the device and models them using two coupled systems: a system of Poisson and drift-diffusion equations describing the nonequilibrium steady state, and a system of Maxwell and drift-diffusion equations describing the transient stage. A ``potential-drop'' boundary condition is enforced on the opposing boundaries of the unit cell to mimic the effect of the bias voltage. Periodic boundary conditions are used for carrier densities and electromagnetic fields. The unit-cell model described by these coupled equations and boundary conditions is discretized using DG methods. Numerical results demonstrate that the proposed DG-based unit-cell scheme has the same accuracy in predicting the THz photocurrent as the DG framework that takes into account the whole device, while it significantly reduces the computational cost.
	\end{abstract}
	
	\section{Introduction}
	Photoconductive devices (PCDs) are promising candidates for terahertz (THz) source generation and signal detection because they are compact and frequency-stable, and capable of operating at room temperature with low optical input power levels~\cite{Lepeshov2017review, Burford2017review, Kang2018review, Yardimci2018review, Yachmenev2019review}. However, the low optical-to-THz conversion efficiency of the conventional PCDs has hindered their widespread use in applications of THz technologies. In recent years, significant progress has been made in alleviating this bottleneck. Introduction of metallic/dielectric nanostructures inside or on top of the PCDs' active region has been shown to increase the conversion efficiency by several orders of magnitude~\cite{Kang2018review, Yardimci2018review, Yachmenev2019review, Yang2014}. This significant increase is attributed to several mechanisms: The optical electromagnetic (EM) field is enhanced due to plasmon~\cite{Yardimci2018review} or Mie~\cite{Yachmenev2019review} resonances; nanostructured electrodes reduce the effective distance that the carriers have to travel~\cite{Berry2013, Yang2014}; and tailoring bias electric field using nanostructures can improve the device efficiency under low carrier densities~\cite{Moon2015}. Furthermore, for both conventional and nanostructured PCDs, physical effects resulting from the coupling between EM fields and carriers alter the device performance. For example, carrier screening causes saturation in the output THz current at high levels of optical pump power~\cite{Darrow1992, Rodriguez1996, Khorshidi2016}. The interplay between all these different physical mechanisms makes the relevant device design very complicated. In this context, rigorous multiphysics simulation tools have become indispensable in understanding these physical mechanisms and their coupling and in enabling and/or accelerating the design process.
	
	Due to their geometrically-intricate structure and the complicated EM wave/field interactions they support, simulation of nanostructured PCDs cannot be carried out using the methods that have been developed for conventional PCDs and rely on semi-analytical approximations/computations~\cite{Sirbu2005, Neshat2010, Kirawanich2008, Khabiri2012, Khiabani2013, Young2014, Moreno2014}. To this end, the finite element method (FEM) has been extensively used in recent years~\cite{Burford2016, Mohammad2016nanoslit, Bashirpour2017, Burford2017}. The optical EM field is calculated with FEM in frequency domain and is used for predicting the carrier generation rate distribution in space. The time dependency of the generation rate is approximated with an analytical expression. Because the frequency-domain solutions are used, the (nonlinear) coupling between carrier dynamics and EM fields is not fully accounted for and consequently the carrier screening effects cannot be modeled by this approach~\cite{Burford2016}. 
	
	Recently, a multiphysics framework making use of discontinuous Galerkin (DG) methods has been proposed~\cite{Chen2019multiphysics, Chen2020steadystate}. This framework solves a coupled system of Poisson and stationary drift-diffusion (DD) equations describing the nonequilibrium steady state of the PCD and a coupled system of time-dependent Maxwell and DD equations describing the transient stage that involves the optoelectronic response of the PCD. The nonlinear coupling between the electrostatic fields and the carriers and that between the EM fields and the carriers are taken into account by the Gummel method and through the use of an explicit time integration scheme, respectively. Even though this DG-based framework provides higher flexibility and higher-order accuracy in both space discretization and time integration and is more robust in modeling nonlinear coupling mechanisms compared to the FEM-based schemes, it is still computationally demanding especially for practical three-dimensional (3D) devices. This is simply because of the multiple space and time characteristic scales involved in the physical processes, e.g., the Debye length $\sim 10~\mathrm{nm}$, the optical wavelength $\sim 100~\mathrm{nm}$, and the device size $\sim 10~\mu\mathrm{m}$~\cite{Chen2019multiphysics}.
	
	One way to reduce this high computational cost is to make use of the nanostructure's periodicity, i.e., model and discretize the multiphysics interactions and their coupling on a unit cell to approximate their behavior on the whole device. This approach calls for proper boundary conditions to be enforced on the surfaces of the unit cell. The periodic boundary conditions (PBCs) required by the optical EM field simulation have been well-studied~\cite{Heshmat2012, Berry2013, Yang2014, Burford2016, Sirenko2018}. However, since a PCD is in a non-equilibrium steady-state under a bias voltage~\cite{Chen2020steadystate, Vasileska2010}, the boundary conditions required by the simulation of the carrier dynamics to be enforced on the unit-cell surfaces are not trivial. In~\cite{Burford2016}, it is assumed that there is no potential-drop along the direction perpendicular to the bias electric field and the optical EM field excitation. A PBC is used along this direction, which reduces the computation domain of the carrier dynamics simulation to a strip containing a chain of unit cells. Even with this approach, the reported computational requirement is high~\cite{Burford2016}. In addition, the nonlinear coupling is not fully considered in~\cite{Burford2016} since a frequency-domain FEM is used to compute the EM field distribution.
	
	In this work, to reduce the high computational cost of nanostructured PCD simulations, a unit-cell scheme is proposed within the multiphysics DG framework developed in~\cite{Chen2019multiphysics, Chen2020steadystate}. For Poisson equation, a ``potential-drop'' boundary condition (PDBC) is enforced on the opposing surfaces of the unit-cell (along the direction of the bias electric field). For carriers and EM fields, PBCs are enforced on the unit-cell surfaces. All boundary conditions are ``weakly'' enforced using the numerical flux of the DG scheme. Numerical results demonstrate that the proposed DG-based unit-cell scheme has practically the same accuracy as the DG framework that takes into account the whole device in predicting the THz photocurrent output. It also retains the main advantages of the DG framework~\cite{Chen2019multiphysics} while significantly reducing the computational cost and making it feasible to simulate practical 3D devices on desktop computers.

	The rest of this paper is organized as follows. In Section~\ref{sec:unitmodel}, the unit-cell model and the relevant boundary conditions are introduced. Sections~\ref{sec:steady} and~\ref{sec:transient} describe the coupled systems of Poisson and stationary DD equations and time-dependent Maxwell and DD equations, respectively. Also, the DG schemes used for discretizing and solving these coupled systems of equations are provided in these sections. In Section~\ref{Validation}, numerical results that validate the accuracy of the proposed scheme and demonstrate its computational benefits are provided. Finally, Section~\ref{Conclusion} concludes the paper and provides some remarks on the future research directions.

	\section{Mathematical model and discretization}
	\label{Scheme}
	\subsection{Unit-cell model}
	\label{sec:unitmodel}
	Fig.~\ref{PM3Dschem} illustrates an example of a 3D nanostructured PCD. The device consists of a photoconductive region (LT-GaAs), a substrate layer (SI-GaAs), two electrodes that are deposited on the photoconductive region, and a metallic nanograting that is placed between them. The grating is designed to support plasmonic modes that enhance the EM fields induced on the structure upon excitation by an optical EM wave~\cite{Yachmenev2019review}.

	The operation of PCDs can be analyzed into two stages. Initially, a bias voltage is applied to the electrodes. The resulting (static) electric field changes the carrier distribution. The re-distributed carriers in turn affects the electric potential distribution. The device reaches a non-equilibrium steady-state mathematically described by a coupled system of Poisson and stationary DD equations~\cite{Vasileska2010}. When an optical EM excitation (i.e., optical pump) is incident on the device, a transient stage starts. The photoconductive material absorbs the EM energy induced on the device due to the optical excitation and generates carriers. The carriers are driven by both the bias electric field and the optical EM fields. The carrier dynamics and EM wave/field interactions are mathematically described by a coupled system of the time-dependent Maxwell and DD equations~\cite{Chen2019multiphysics, Chen2019discontinuous}.
	
	To accurately capture these coupled physical processes that occur on the whole device using only a single unit cell [as illustrated in Fig.~\ref{PM3Dschem} (b)], appropriate boundary conditions must be enforced on the unit-cell surfaces. 
	
	For Poisson equation, one can not simply use PBCs since the potential drops from the anode to the cathode. A critical observation that makes modeling the biased state using a unit cell possible is, in the electrostatic problem: The nanostructure generates only local variations in the potential~\cite{Chen2020float, Chen2020hybridizable} and on average the potential drops linearly between the two electrodes (as in the homogeneous case without the nanostructure)~\cite{Menshov2014}. Furthermore, since all unit cells are the same, the potential drop and the local potential variation within each unit cell should be the same. This analysis suggests that the bias static electric field, which is equal to the gradient of the potential, is periodic. Therefore, the steady-state carrier densities and the field dependent mobility should also be periodic. Furthermore, since the nanostructures and the carrier distributions are periodic, the optical EM fields induced on the structure are the same in all unit cells. The same argument applies to carrier dynamics since the mobility, the static electric field, and the optical EM fields are the same in all unit cells. Therefore, PBCs can be used for stationary DD equations, time-dependent Maxwell equations, and time-dependent DD equations.
	
	The boundary conditions discussed above are mathematically described as follows. For Poisson equation, the boundary conditions are
	\begin{align}
	\label{PDBC} & { \varphi } (-w_x/2,y,z) = { \varphi }(w_x/2,y,z) + {\varphi _{\mathrm{drop}}}(y,z), \\
	\label{PBC}  & { \varphi } (x,-w_y/2,z) = { \varphi }(x,w_y/2,z),
	\end{align}
	where $\varphi(x,y,z)$ is the electric potential, $w_x$ and $w_y$ are the widths of the unit cell in $x$ and $y$ directions, respectively, and $\varphi _{\mathrm{drop}}(y,z)$ is the potential-drop between the two faces of the unit cell perpendicular to the $x$ direction. In the rest of the text, \eqref{PDBC} is termed as PDBC. Note that the PBC~\eqref{PBC} is used along the $y$ direction because the potential does not drop along this direction (hence is periodic)~\cite{Burford2016}. The potential drop function ${\varphi _{\mathrm{drop}}}(y,z)$ in \eqref{PDBC} is selected as described next. 
	${\varphi _{\mathrm{drop}}}(y,z)$ is position-dependent, e.g., the potential drop becomes smaller away from the electrodes in the $-z$ direction. Since the height of the device is much smaller than its width~\cite{Lepeshov2017review, Burford2017review, Kang2018review} and the electrodes extend to the whole width of the device, the potential drop is approximately the same for all $y$ and $z$ at a given value of $x$ (i.e., on a surface perpendicular to $x$ direction). Therefore, ${\varphi _{\mathrm{drop}}}(y,z)$ can be simplified to a single value ${\varphi _{\mathrm{drop}}}$. Additionally, as discussed before, on average the potential drops linearly between the two electrodes~\cite{Menshov2014, Chen2019unitcell}. Consequently, $\varphi_{\mathrm{drop}}(y,z)$ can be estimated as: ${\varphi _{\mathrm{drop}}}(y,z) = {w_{x}}({{V_{\mathrm{bias}}}}\left / w_{\mathrm{sd}} \right. )$, where the ${V_{\mathrm{bias}}}$ is the bias voltage applied to the electrodes and $w_{\mathrm{sd}}$ is the distance between them. 
	
	For stationary DD equations and time-dependent DD and Maxwell equations, PBCs are used:
	\begin{align}
	{ U }(-{w_x}/2,y,z) = { U }({w_x}/2,y,z), \\
	{ U }(x,-{w_y}/2,z) = { U }(x,{w_y}/2,z),
	\end{align}
	where $U(x,y,z)$ represents the steady-state electron/hole density, the transient electron/hole density, or the transient electric/magnetic field. For all equations, the boundary conditions on the top and bottom surfaces of the unit cell (surfaces perpendicular to $z$ direction) are the same as those would be used in the full-device simulation~\cite{Chen2020steadystate, Chen2019multiphysics}.
	
	Two comments about the unit-cell model introduced in this section are in order: (i) The approximation of using a single value for potential drop can be improved by estimating ${\varphi _{\mathrm{drop}}}(y,z)$ from the solution of the same device but without the nanostructure (which generates only local variations in the potential). Modeling a simpler device without the nanostructure is easier since a much coarser mesh can be used. (ii) The nanostructure does not have to be metallic for the unit-cell model to hold; it is also applicable when the nanostructure is made of a dielectric material~\cite{Lepeshov2017review, Burford2017review, Kang2018review, Yardimci2018review, Yachmenev2019review, Fang2018}.
	
	
	\subsection{Unit-cell Poisson-DD solver}
	\label{sec:steady}
	The coupled system of Poisson and stationary DD equations is solved using the Gummel iteration method~\cite{Chen2020steadystate, Vasileska2010}. In each iteration, three partial differential equations (PDEs), namely the linearized Poisson equation and two DD equations are solved~\cite{Chen2020steadystate}. These equations, together with the boundary conditions described above, are discretized using DG methods. 
	
	The Poisson equation in the unit cell is expressed as a boundary value problem (BVP)
	\begin{alignat}{2}
	\label{BVP_PS0}
	& \nabla  \cdot [\varepsilon ({\mathbf{r}}){\mathbf{E}}({\mathbf{r}})] + g({\mathbf{r}})\varphi ({\mathbf{r}}) = f({\mathbf{r}}), \quad && {\mathbf{r}} \in \Omega, \\
	\label{BVP_PS1}
	& {\mathbf{E}}({\mathbf{r}}) =  - \nabla \varphi ({\mathbf{r}}), && {\mathbf{r}} \in \Omega, \\
	\label{BVP_PS4} 
	& { \varphi }(-{w_x}/2,y,z) \! = \! { { \varphi } }({w_x}/2,y,z) \! + \! {\varphi _{\mathrm{drop}}}(y,z), \hspace{0.12cm} \quad && {\mathbf{r}} \in \partial {\Omega _x}, \\
	\label{BVP_PS5} & { \varphi }(x,-{w_y}/2,z) \! = \! { \varphi }(x,{w_y}/2,z),  && {\mathbf{r}} \in \partial {\Omega _y}, \\
	\label{BVP_PS3} & {\mathbf{\hat n}}({\mathbf{r}}) \cdot \varepsilon ({\mathbf{r}}){\mathbf{E}}({\mathbf{r}}) \! = \! {f_N}({\mathbf{r}}), && {\mathbf{r}} \in \partial {\Omega _z},
	\end{alignat}
	where $\varphi ({\mathbf{r}})$ and ${\mathbf{E}}({\mathbf{r}})$ are the unknowns to be solved for, $g(\textbf{r})$ and $f(\textbf{r})$ are known coefficients, $\varepsilon(\textbf{r})$ is the permittivity, $\Omega$ is the solution domain, $\partial {\Omega _\nu}$ represents the surfaces perpendicular to the $\nu$-direction, $\nu\in\{x,y,z\}$, ${\mathbf{\hat n}}({\mathbf{r}})$ is the outward pointing normal vector on $\partial {\Omega _\nu}$. The homogeneous Neumann boundary condition $f_N(\mathbf{r})=0$ is used in~\eqref{BVP_PS3}~\cite{Chen2019multiphysics}.

	Discretizing $\Omega$ into a set of non-overlapping elements, DG approximates the unknowns with basis functions (the nodal basis function~\cite{Hesthaven2008} is used in this work) in each element and applies Galerkin testing to the resulting equations~\cite{Cockburn1998, Hesthaven2008, Shu2016}. This process yields a matrix system as
	\begin{equation}
	\label{LS_PS0}
	\left[ {\begin{array}{*{20}{c}}
		{{{\bar M}^g}}&{\bar D\bar \varepsilon } \\ 
		{\bar G}&{{{\bar M}^{\mathbf{E}}}} 
		\end{array}} \right]\left[ \begin{gathered}
	{\bar \Phi } \hfill \\
	{\bar E} \hfill \\ 
	\end{gathered}  \right] = \left[ \begin{gathered}
	{{\bar B}^\varphi } \hfill \\
	{{\bar B}^{\mathbf{E}}} \hfill \\ 
	\end{gathered}  \right].
	\end{equation}
	Here, $\bar \Phi$ and $\bar E$ are unknown vectors storing the basis expansion coefficients, ${{\bar M}^g}$ and ${\bar M}^{\mathbf{E}}$ are block diagonal mass matrices, $\bar G$ and $\bar D$ are block sparse matrices representing the gradient and divergence operators, respectively, and $\bar B^\varphi$ and $\bar B^{\mathbf{E}}$ have contributions from the tested force term and boundary conditions. Detailed expressions of these vectors and matrices can be found in~\cite{Chen2020steadystate}.

	The continuity of solutions across element interfaces is enforced through a uniquely defined numerical flux. For Poisson equation, the alternate flux~\cite{Cockburn1998}
	\begin{align}
	\label{phiflux} & {\varphi ^*} = \left\{ \varphi  \right\} + 0.5\boldsymbol{\hat \beta} \cdot {\mathbf{\hat n}}\left[\kern-0.15em\left[ \varphi  
	\right]\kern-0.15em\right], \\
	\label{Eflux} & {\left( {\varepsilon {\mathbf{E}}} \right)^*} = \left\{ {\varepsilon {\mathbf{E}}} \right\} - 0.5\boldsymbol{\hat \beta} ({\mathbf{\hat n}} \cdot \left[\kern-0.15em\left[ {\varepsilon {\mathbf{E}}} 
	\right]\kern-0.15em\right])
	\end{align}
	is used on each element surface in the interior of $\Omega$, where ``average'' and ``jump'' operators are defined as $\left\{\odot \right\} = 0.5({\odot ^ - } + { \odot ^ + })$ and $\left[\kern-0.15em\left[  \odot  
	\right]\kern-0.15em\right] = { \odot ^ - } - { \odot ^ + }$, respectively, with $ \odot $ being a scalar or a vector variable. The same definitions of these operators are used throughout this paper. Superscripts ``-'' and ``+'' refer to variables defined in the interior and exterior of the surface, respectively. $\boldsymbol{\hat \beta}={\mathbf{\hat n}}$ determines the upwind direction of $\varphi $ and $(\varepsilon {\mathbf{E}})$~\cite{Cockburn1998, Hesthaven2008, Shu2016}. On $\partial {\Omega _N}$, the numerical fluxes are chosen as ${\varphi ^*} = {\varphi ^ - }$ and ${\left( {\varepsilon {\mathbf{E}}} \right)^*} = {f_N}$~\cite{Cockburn1998}. Here, the variables are defined on the surfaces and the explicit dependency on ${\mathbf{r}}$ is dropped for the simplicity of the notation.
	
	To enforce PBC, same meshes are created on the opposing surfaces of the unit cell and each element face on a given surface is ``connected'' to its counterpart on the opposing surface. This means that when calculating the numerical flux on the boundary, \eqref{phiflux}-\eqref{Eflux} are used but, for each element face, the exterior variable is taken from its ``connected'' face on the opposing surface
	\begin{align}
	\label{PBC_PS0} & U^+(x,-w_y/2,z)  = {U(x,w_y/2,z)},\\
	\label{PBC_PS1} & U^+(x,w_y/2,z) = {U(x,-w_y/2,z)},
	\end{align}
	where $U \in \{\varphi, (\varepsilon\mathbf{E})\}$.
	
	For PDBC, the element faces on the opposing surfaces are ``connected'' just like it is done for the PBC. {\color{black} An intermediate state is defined as $\varphi^* = {\varphi(w_x/2,y,z)} + 0.5{\varphi _{\mathrm{drop}}}(y,z) = {\varphi(-w_x/2,y,z)} - 0.5{\varphi _{\mathrm{drop}}}(y,z)$, and} the exterior variables in the numerical flux expressions are set as
	\begin{align}
	& \varphi^+(-w_x/2,y,z)  = \varphi^* ,\\
	& \varphi^+(w_x/2,y,z)  = \varphi^* ,\\
	& (\varepsilon\mathbf{E})^+(-w_x/2,y,z) = (\varepsilon\mathbf{E})(w_x/2,y,z) ,\\
	& (\varepsilon\mathbf{E})^+(w_x/2,y,z)  = (\varepsilon\mathbf{E})(-w_x/2,y,z).
	\end{align}

	The electron DD equations in the unit cell are expressed as a BVP~\cite{Chen2020steadystate}
	\begin{alignat}{2}
	\label{BVP_DD0}
	& \nabla  \cdot [d({\mathbf{r}}){\mathbf{q}}({\mathbf{r}})]{\text{ + }}\nabla  \cdot [{\mathbf{v}}({\mathbf{r}})n({\mathbf{r}})] = R({\mathbf{r}}), \quad  && {\mathbf{r}} \in \Omega ,\\
	\label{BVP_DD1}
	& {\mathbf{q}}({\mathbf{r}}) = \nabla n({\mathbf{r}}), && {\mathbf{r}} \in \Omega ,\\
	\label{BVP_DD4}
	& { n }(-{w_x}/2,y,z) \! = \! { { n } }({w_x}/2,y,z) \!  && {\mathbf{r}} \in \partial {\Omega _x} ,\\
	\label{BVP_DD5} 
	& { n }(x,-{w_y}/2,z) \! = \! { n }(x,{w_y}/2,z),  && {\mathbf{r}} \in \partial {\Omega _y} ,\\
	\label{BVP_DD3}
	& {\mathbf{\hat n}}({\mathbf{r}}) \cdot [d({\mathbf{r}}){\mathbf{q}}({\mathbf{r}}) + {\mathbf{v}}({\mathbf{r}})n({\mathbf{r}})] \! = \! {f_R}({\mathbf{r}}), && {\mathbf{r}} \in \partial {\Omega _z},
	\end{alignat}
	where $n({\mathbf{r}})$ and ${\mathbf{q}}({\mathbf{r}})$ are the unknowns to be solved for, and $d({\mathbf{r}})$ ${\mathbf{v}}({\mathbf{r}})$, and $R({\mathbf{r}})$ are known coefficients within the Gummel method~\cite{Chen2020steadystate}. 
	The homogeneous Robin boundary condition $f_R({\mathbf{r}})=0$ is used in~\eqref{BVP_DD3}~\cite{Chen2019multiphysics}. The BVP for holes only differs by the sign in front of the drift term.
	

	DG discretization of~\eqref{BVP_DD0}-\eqref{BVP_DD3} yields a matrix system as
	\begin{equation}
	\label{LS_DD0}
	\left[ {\begin{array}{*{20}{c}}
		{\bar C}&{\bar D\bar d} \\ 
		{ - \bar G}&{{{\bar M}^{\mathbf{q}}}} 
		\end{array}} \right]\left[ \begin{gathered}
	{\bar N} \hfill \\
	{\bar Q} \hfill \\ 
	\end{gathered}  \right] = \left[ \begin{gathered}
	{{\bar B}^n} \hfill \\
	{{\bar B}^{\mathbf{q}}} \hfill \\ 
	\end{gathered}  \right],
	\end{equation}
	where $\bar N$ and $\bar Q$ are unknown vectors storing the basis expansion coefficients, ${\bar M^{\mathbf{q}}}$ is same as ${\bar M^{\mathbf{E}}}$, $\bar G$ and $\bar D$ are same as before, $\bar d$ is a diagonal matrix, and $\bar C$ corresponds to the drift term~\cite{Chen2020steadystate}.

	The local Lax-Friedrichs flux~\cite{Chen2020steadystate}
	\begin{equation*}
	\label{fluxDD2} {\left( {{\mathbf{v}}n} \right)^*} \! = \! \left\{ {{\mathbf{v}}n} \right\} + \alpha {\mathbf{\hat n}}({n^- } - {n^ + }), \alpha \! = \! \max (|{\mathbf{\hat n}} \cdot {{\mathbf{v}}^ - }|,|{\mathbf{\hat n}} \cdot {{\mathbf{v}}^ + }|)/2
	\end{equation*}
	is used for the drift term and the alternate flux \eqref{phiflux}-\eqref{Eflux} is used for the diffusion term ($\varphi$ and $(\varepsilon \mathbf{E})$ are replaced with $n$ and $\mathbf{q}$, respectively).
	
	The PBC~\eqref{BVP_DD5} is enforced just like it is done in~\eqref{PBC_PS0}-\eqref{PBC_PS1}, with $U \in \{n, (d\mathbf{q})\}$, and for \eqref{BVP_DD4}
	\begin{align}
	\label{PBC_DD0} & U^+(-w_x/2,y,z)  = {U(w_x/2,y,z)}, \\
	\label{PBC_DD1} & U^+(w_x/2,y,z) = {U(-w_x/2,y,z)}.
	\end{align}

	Solutions of the matrix systems~\eqref{LS_PS0} and~\eqref{LS_DD0} are computed using linear solvers at every iteration of the Gummel method. The steady-state solutions are obtained after the Gummel iteration converges and are used as inputs for the transient solver~\cite{Chen2019multiphysics}.
	
	
	
	\subsection{Unit-cell Maxwell-DD solver}
	\label{sec:transient}
	The coupled system of time-dependent Maxwell and DD equations is integrated in time using an explicit scheme. The nonlinear coupling between the Maxwell and DD equations is accounted for by alternately feeding their updated solutions into each other during the time integration~\cite{Chen2019multiphysics}. Each set of equations is discretized using a time-domain DG (DGTD) method~\cite{Cockburn1998, Hesthaven2008, Shu2016} that uses its own time-step size~\cite{Chen2019multiphysics}.
	
	The time-dependent electron DD equations in the unit cell are expressed as an initial-boundary value problem (IBVP)~\cite{Chen2019multiphysics}
	\begin{alignat}{2}
	\label{IVP_DD0}
	& {\partial _t}n({\mathbf{r}},t) = \nabla \cdot [d({\mathbf{r}}){\mathbf{q}}({\mathbf{r}},t)] +
	\nabla \cdot [{\mathbf{v}}({\mathbf{r}},t)n({\mathbf{r}},t)] - R({\mathbf{r}},t), \quad && {\mathbf{r}} \in \Omega ,\\
	\label{IVP_DD1}
	& {\mathbf{q}}({\mathbf{r}},t) = \nabla n({\mathbf{r}},t), && {\mathbf{r}} \in \Omega ,\\
	\label{IVP_DD4}
	& { n }(-{w_x}/2,y,z,t) \! = \! { { n } }({w_x}/2,y,z,t) \!  && {\mathbf{r}} \in \partial {\Omega _x} ,\\
	\label{IVP_DD5} 
	& { n }(x,-{w_y}/2,z,t) \! = \! { n }(x,{w_y}/2,z,t),  && {\mathbf{r}} \in \partial {\Omega _y} ,\\
	\label{IVP_DD3}
	& \mathbf{\hat{n}}({\mathbf{r}}) \cdot [d({\mathbf{r}}) {\mathbf{q}} ({\mathbf{r}},t) + {\mathbf{v}}({\mathbf{r}},t)n({\mathbf{r}},t)] = {f_R}({\mathbf{r}}), \hspace{0.2cm} && {\mathbf{r}} \in \partial {\Omega _z} ,
	\end{alignat}
	where $n({\mathbf{r}},t)$ and ${\mathbf{q}}({\mathbf{r}},t)$ are the unknowns to be solved for,
	$d({\mathbf{r}})$ and ${\mathbf{v}}({\mathbf{r}},t)$ are known coefficients, $R({\mathbf{r}},t) = {R^t}(n_e^t,n_h^t) - G({{\mathbf{E}}^t},{{\mathbf{H}}^t})$ is the net-recombination, $R^t({n_e},{n_h})$ is the transient recombination rate, and $G({\mathbf{E}},{\mathbf{H}})$ is the generation rate~\cite{Moreno2014, Chen2019multiphysics}. Here, to simplify the notation, the subscript ``e'' is dropped. The IBVP for holes only differs by the sign in front of the drift term.

	Apart from the time dependency and the time derivative term on the left side of \eqref{IVP_DD0}, the system has the same form as \eqref{BVP_DD0}-\eqref{BVP_DD3}. Using the same spatial discretization as that used for \eqref{BVP_DD0}-\eqref{BVP_DD3}, one can obtain the semi-discrete form~\cite{Chen2019multiphysics}
	\begin{align}
	\label{semiDD0} &
	{{\bar M}_{kk}}{\partial _t}{{\bar N}_k(t)} = {{\bar C}_{kk}}{{\bar N}_k(t)} + {{\bar C}_{kk'}}{{\bar N}_{k'}(t)} + { {\bar D}_{kk} {\bar d}_k }{{\bar Q}_k(t)} + { {\bar D}_{kk'} {\bar d}_{k'} }{{\bar Q}_{k'}(t)} - \bar B_k^n(t) ,\\
	\label{semiDD1} &
	\bar M_{kk}^{\mathbf{q}}{{\bar Q}_k(t)} = {{\bar G}_{kk}}{{\bar N}_k}(t) + {{\bar G}_{kk'}}{{\bar N}_{k'}}(t) + \bar B_k^{\mathbf{q}} ,
	\end{align}
	where ${\bar N_k(t)}$ and ${\bar Q}_k(t)$ unknown vectors storing time-dependent basis expansion coefficients, $\bar B_k^n(t)$ and $\bar B_k^{\mathbf{q}}$ are vectors constructed from the net-recombination and boundary conditions (other than PBC), ${{\bar M}_{kk}}$, ${\bar C_{kk}}$/${\bar C_{kk'}}$, ${\bar G_{kk}}$/${\bar G_{kk'}}$, and ${{\bar D}_{kk}}$/${{\bar D}_{kk'}}$ are the elemental mass, convection, gradient, and divergence matrices, respectively~\cite{Chen2019multiphysics}.
	
	The numerical fluxes and boundary conditions are the same as those used in the stationary DD equations. The semi-discretized system~\eqref{semiDD0}-\eqref{semiDD1} is integrated in time using an explicit third-order total-variation-diminishing Runge-Kutta scheme~\cite{Shu1988}.

	Likewise, Maxwell equations in the unit cell are expressed as the following IBVP
	\begin{alignat}{2}
	\label{IVP_Max0} & \varepsilon(\mathbf{r}) {\partial _t}{{\mathbf{E}}}({\mathbf{r}},t) =  \nabla \times {{\mathbf{H}}}({\mathbf{r}},t) - {\mathbf{J}}({\mathbf{r}},t), \quad && {\mathbf{r}} \in \Omega ,\\
	\label{IVP_Max1} & \mu(\mathbf{r})  {\partial _t}{{\mathbf{H}}}({\mathbf{r}},t) = -\nabla \times {{\mathbf{E}}} ({\mathbf{r}},t), && {\mathbf{r}} \in \Omega ,\\
	\label{IVP_Max2}
	& { U }(-{w_x}/2,y,z,t) \! = \! { { U } }({w_x}/2,y,z,t) \!  && {\mathbf{r}} \in \partial {\Omega _x} ,\\
	\label{IVP_Max3} 
	& { U }(x,-{w_y}/2,z,t) \! = \! { U }(x,{w_y}/2,z,t),  && {\mathbf{r}} \in \partial {\Omega _y} ,\\
	\label{IVP_Max4}
	& \nabla \times \mathbf{E}(\mathbf{r},t) = 0, \nabla \cdot \mathbf{H}(\mathbf{r},t) = 0, && {\mathbf{r}} \in \partial {\Omega _z} ,
	\end{alignat}
	where ${{\mathbf{E}}}({\mathbf{r}},t)$ and ${{\mathbf{H}}}({\mathbf{r}},t)$ are the transient electric and magnetic fields, $\mu(\textbf{r})$ is the permeability, ${\mathbf{J}}({\mathbf{r}},t)$ is the transient current density, and $U \in \{\mathbf{E}, \mathbf{H}\}$. Discretizing \eqref{IVP_Max0}-\eqref{IVP_Max4} with the DGTD method~\cite{Hesthaven2008, Sirenko2012, Li2015IBC, Li2018Q2D, Lu2004} yields
	\begin{align}
	\label{semiMaxwell0}
	& {\varepsilon _k}{\partial _t}\bar E_k^x(t) \! = \! \bar D_k^y\bar H_k^z(t) \! - \! \bar D_k^z\bar H_k^y(t) \! - \! {\bar{F}_k}F_k^{{\mathbf{E}},x}(t) \! - \! \bar J_k^x(t) ,\\
	\label{semiMaxwell1}
	& {\varepsilon _k}{\partial _t}\bar E_k^y(t) \! = \! \bar D_k^z\bar H_k^x(t) \! - \! \bar D_k^x\bar H_k^z(t) \! - \! {\bar{F}_k}F_k^{{\mathbf{E}},y}(t) \! - \! \bar J_k^y(t) ,\\
	\label{semiMaxwell2}
	& {\varepsilon _k}{\partial _t}\bar E_k^z(t) \! = \! \bar D_k^x\bar H_k^y(t) \! - \! \bar D_k^y\bar H_k^x(t) \! - \! {\bar{F}_k}F_k^{{\mathbf{E}},z}(t) \! - \! \bar J_k^z(t) ,\\
	\label{semiMaxwell3}
	& {\mu _k}{\partial _t}\bar H_k^x(t) \! = \! - [\bar D_k^y\bar E_k^z(t) \! - \! \bar D_k^z\bar E_k^y(t)] \! + \! {\bar{F}_k}F_k^{{\mathbf{H}},x}(t) ,\\
	\label{semiMaxwell4}
	& {\mu _k}{\partial _t}\bar H_k^y(t) \! = \! - [\bar D_k^z\bar E_k^x(t) \! - \! \bar D_k^x\bar E_k^z(t)] \! + \! {\bar{F}_k}F_k^{{\mathbf{H}},y}(t) ,\\
	\label{semiMaxwell5}
	& {\mu _k}{\partial _t}\bar H_k^z(t) \! = \! - [\bar D_k^x\bar E_k^y(t) \! - \! \bar D_k^y\bar E_k^x(t)] \! + \! {\bar{F}_k}F_k^{{\mathbf{H}},z}(t) ,
	\end{align}
	where $\bar E_k^\nu(t)$ and $\bar H_k^\nu(t)$ are unknown vectors storing time-dependent basis expansion coefficients, $\bar J_k^\nu(t)$ is the current density vector, $\varepsilon_k$ and $\mu_k$ are the constant permittivity and permeability in element $k$, $\bar D_k^\nu = {\bar{M}_k^{-1}} \tilde S_k^\nu $, and $\bar{F}_k = {\bar{M}_k^{-1}} {\tilde L_k}$, where ${\bar M_k}$, {\color{black}${\tilde S_k^\nu}$}, and ${\tilde L_k}$ are the mass, stiffness, and surface mass matrices, respectively~\cite{Hesthaven2008}.
	
	In~\eqref{semiMaxwell0}-\eqref{semiMaxwell5}, $F_k^{{\mathbf{E}},\nu }(t)$ and $F_k^{{\mathbf{H}},\nu }(t)$ are the $\nu$ components of $\hat{\mathbf{n}} \times [{\mathbf{H}_k}(t) - {\mathbf{H}}_k^*(t)]$ and $\hat{\mathbf{n}} \times [{\mathbf{E}_k}(t) - {\mathbf{E}}_k^*(t)]$, respectively, $\nu \in \{x,y,z\}$. 
	The upwind flux~\cite{Hesthaven2008} is used here:
	\begin{align}
	&  {{\mathbf{E}}^*} = (2\left\{ {Y{\mathbf{E}}} \right\} - {\mathbf{\hat n}} \times \left[\kern-0.15em\left[ {\mathbf{H}} 
	\right]\kern-0.15em\right])/(2\left\{ Y \right\}) ,\\
	&  {{\mathbf{H}}^*} = (2\left\{ {Z{\mathbf{H}}} \right\} + {\mathbf{\hat n}} \times \left[\kern-0.15em\left[ {\mathbf{E}} 
	\right]\kern-0.15em\right])/(2\left\{ Z \right\}) ,
	\end{align}
	where $Z$ and $Y$ are the wave impedance and wave admittance, respectively. The PBCs~\eqref{IVP_Max2} and~\eqref{IVP_Max3} are enforced similarly as~\eqref{PBC_PS0}-\eqref{PBC_PS1} and~\eqref{PBC_DD0}-\eqref{PBC_DD1}, with $U \in \{\mathbf{E}, \mathbf{H} \}$. The perfect electric conductor (PEC) boundary condition is used on $\partial {\Omega _z}$, ${\mathbf{\hat n}} \times \left[\kern-0.15em\left[ {\mathbf{E}} \right]\kern-0.15em\right] = 2{\mathbf{\hat n}} \times {{\mathbf{E}}^ - }$, ${\mathbf{\hat n}} \times \left[\kern-0.15em\left[ {\mathbf{H}} \right]\kern-0.15em\right] = 0$ and perfectly matched layers (PMLs)~\cite{Berenger1994, Lu2004, Gedney2009, Chen2020pml} are used for absorbing outgoing EM waves. The semi-discrete system~\eqref{semiMaxwell0}-\eqref{semiMaxwell5} is integrated in time using a five-stage fourth-order Runge-Kutta method~\cite{Hesthaven2008}.

	\section{Numerical results}
	\label{Validation}

	\begin{table}[!b]
		\centering
		\begin{threeparttable}
			\renewcommand{\arraystretch}{1.2}
			\caption{Semiconductor material parameters used in the PCD example}
			\label{parameters}
			\begin{tabular}{c c}
				\hline
				C & $1.3\times 10^{16}$ cm$^{-3}$ \\
				$n_i$ & $9\times 10^6$ cm$^{-3}$ \\ 
				\hline
				Mobility &
				{ \begin{math} \begin{array}{cc}
					\mu_e^0=8000 \mathrm{cm}^2\mathrm{/V/s}, \mu_h^0=400 \mathrm{cm}^2\mathrm{/V/s} \\
					V_e^{sat}\!=\!1.725 \! \times \! 10^{7} \mathrm{cm/s}, V_h^{sat} \! = \! 0.9 \! \times \! 10^{7}\mathrm{cm/s} \\
					\beta_e=1.82, \beta_h=1.75
					\end{array} \end{math} } \\ \hline
				Recombination & 
				{ \begin{math} \begin{array}{cc}
					\tau_e=0.3 \mathrm{ps}, \tau_h=0.4 \mathrm{ps} \\
					n_{e1}=n_{h1}=4.5 \times 10^6 \mathrm{cm}^{-3} \\
					C_e^A=C_h^A=7\times 10^{-30} \mathrm{cm}^6\mathrm{/s}
					\end{array} \end{math} } \\
				\hline
			\end{tabular}
		\end{threeparttable}
	\end{table}
	
\begin{figure}[!b]
		\captionsetup[subfigure]{position=top, labelfont=bf,textfont=normalfont,singlelinecheck=off,justification=raggedright}
		\centering
		\subfloat[\label{Fig1a}]{\includegraphics[height=6cm]{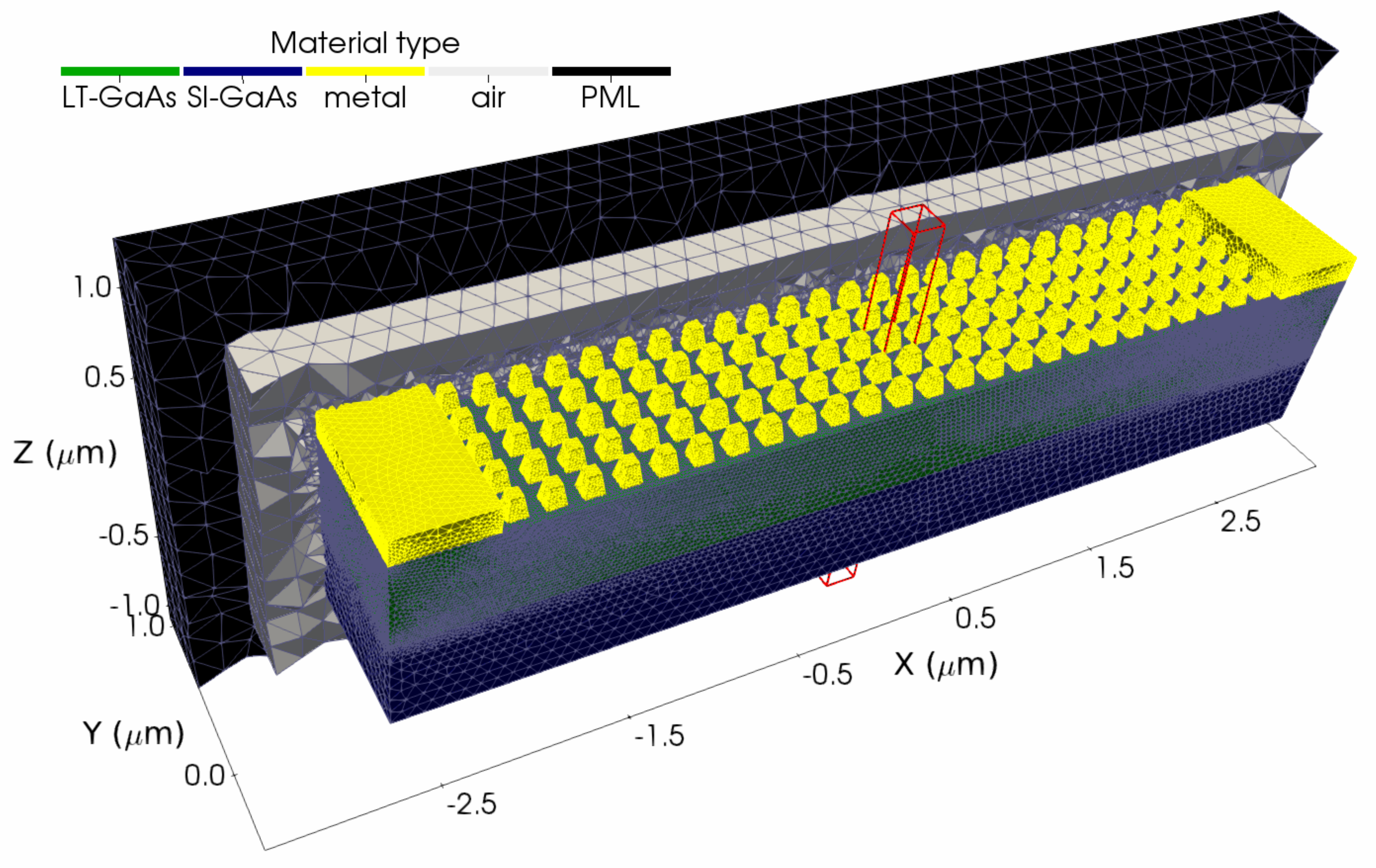}} \vspace{0.1cm}
		\subfloat[\label{Fig1b}]{\includegraphics[height=6cm]{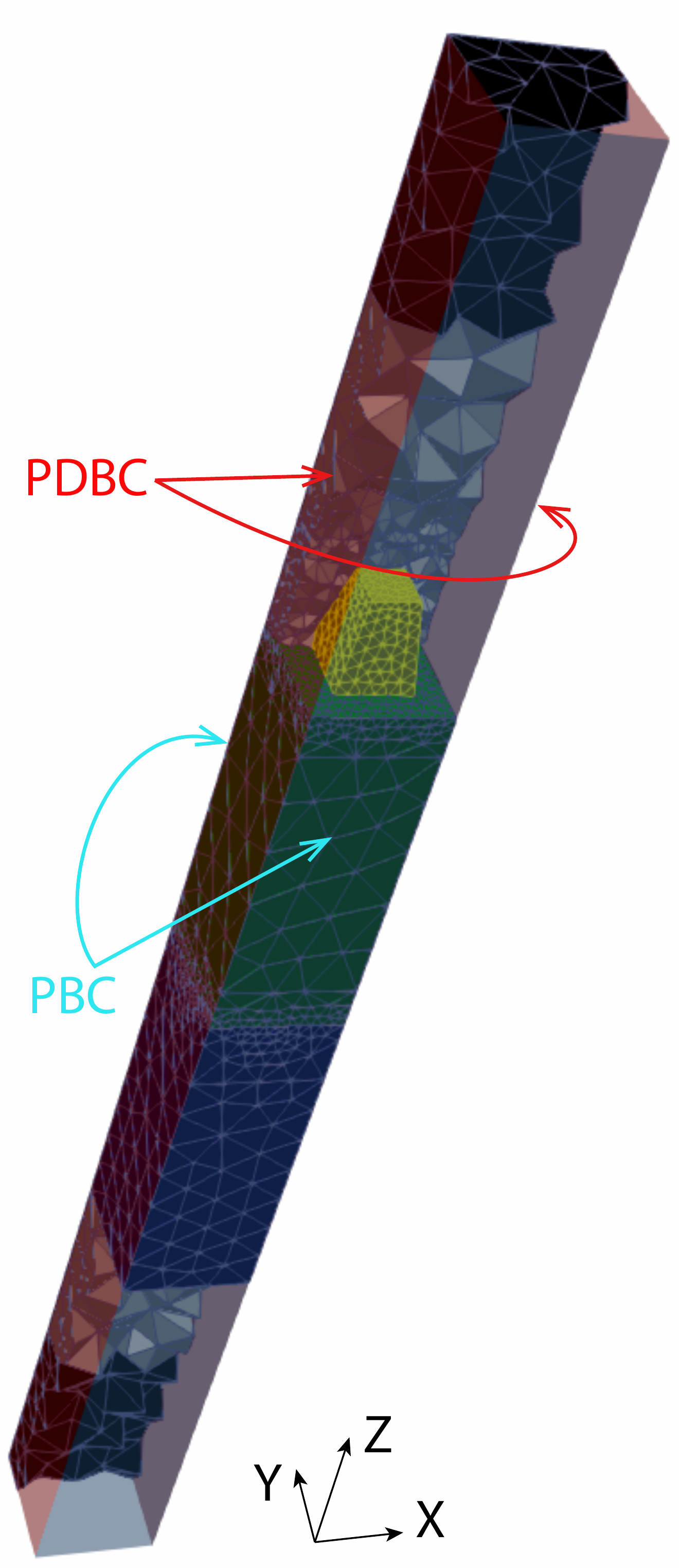}}
		\caption{(a) Geometry of a nanostructured PCD and the corresponding mesh used in the full-device DG scheme. (b) The unit cell used by the proposed scheme for the PCD geometry in (a).}
		\label{PM3Dschem}
	\end{figure}

	To validate the proposed scheme, a 3D nanostructured PCD is simulated using the proposed method (with only a unit cell) and the results are compared to those obtained for the full device using the DG framework~\cite{Chen2020steadystate, Chen2019multiphysics} (which has been validated against experimental data). The device is illustrated in Fig.~\ref{PM3Dschem}. The thickness of the LT-GaAs and the SI-GaAs layers is $0.5~\mu$m. The nanograting has a periodicity of $0.18~\mu$m in $x$ and $y$ directions and its height is $0.12~\mu$m. The metal block is a truncated square pyramid with dimensions of $0.06~\mu$m and $0.1~\mu$m on its top and bottom, respectively. The semiconductor material parameters are same as those used in~\cite{Chen2019multiphysics} and are provided in Table~\ref{parameters}. The permittivity of LT-GaAs is modeled using the Lorentz dispersion relation
	\begin{equation*}
	\varepsilon(\omega) = \varepsilon_{0} (\varepsilon_{\infty} + {\frac{\omega_{p}^2}{ \omega_{o}^2 - \omega^2 - i \gamma \omega } })
	\end{equation*}
	with ${\color{black}\varepsilon}_{\infty}=5.785$, $\omega_o=4.783\times10^{15}~\mathrm{rad/s}$, $\gamma=4.557\times10^{14}~\mathrm{rad/s}$, $\omega_p=1.061\times10^{16}~\mathrm{rad/s}$. SI-GaAs is modeled as a dielectric material with ${\color{black}\varepsilon}_\mathrm{r}=13.26$. The nanogratings and contacts are modeled as gold using the Drude model
	\begin{equation*}
	\varepsilon(\omega) = \varepsilon_{0} (\varepsilon_{\infty} - {\frac{\omega_{p}^2}{ \omega^2 + i \gamma \omega } })
	\end{equation*}
	with parameters ${\color{black}\varepsilon}_{\infty}=1.0$, $\omega_p=1.372\times10^{16}~\mathrm{rad/s}$, $\gamma=8.052\times10^{13}~\mathrm{rad/s}$~\cite{Chen2019multiphysics}. All materials are considered nonmagnetic $\mu_\mathrm{r}=1.0$. The DD equations are solved only within the LT-GaAs layer while Poisson equation and the time-dependent Maxwell equations are solved everywhere in the unit cell (and the full device to obtain the reference results). {\color{black}The bias voltage is fixed at $V_{\mathrm{bias}}=10$ V. The PCD is operated in the continuous-wave mode~\cite{Lepeshov2017review, Yardimci2018review} and excited from top by two continuous-wave lasers operating at $374.5~\mathrm{THz}$ and $375.5~\mathrm{THz}$, with a frequency difference of $1~\mathrm{THz}$.} The laser beam width is $1.8~\mu\mathrm{m}$ in the full-device simulation.

	\begin{table}[!b]
		\centering
		\begin{threeparttable}
			\renewcommand{\arraystretch}{1.1}
			\caption{Computational time\tnote{*} required by the full-device and unit-cell simulations.}
			\label{cost}
			\setlength{\tabcolsep}{3pt}
			\vspace{-0.2cm}
			\begin{tabular}{   p{60pt} | p{26pt}  p{26pt}  p{44pt} | p{26pt}  p{26pt}  p{44pt} }
				\hline 
				& \multicolumn{3}{|c|}{steady state} & \multicolumn{3}{|c}{transient stage} \\ 
				\hline
				& core & hour & core-hour & core & hour & core-hour\\ \hline
				full device & 32 & 201 & 6432 & 4096 & 35 & 143360 \\ \hline
				unit cell & 16 & 0.22 & 3.52 & 32 & 3 & 96 \\ \hline
			\end{tabular}
			\smallskip
			\scriptsize
			\begin{tablenotes}
				\item[*] {Tested on Shaheen II (https://www.hpc.kaust.edu.sa/), a Cray XC40 system consists of 6174 nodes based on Intel Xeon (R) E5-2698 v3.}
			\end{tablenotes}
		\end{threeparttable}
	\end{table}

	For the unit-cell simulation, the domain is shown in Fig.~\ref{PM3Dschem} (b) and the boundary conditions are those explained in Section~\ref{Scheme}. For the full-device simulation, the height and width of the device (in $x$ and $y$ directions, respectively) are $3.1~\mu$m and $0.54~\mu$m. The nanograting is made of a $15 \times 3$ grid of unit cells, which is smaller than practical devices but is large enough for the purpose of validation in this work. For the DD equations, the Dirichlet boundary condition ${n_e} = (C + \sqrt {{C^2} + 4{n_i^2}} )/2$, ${n_h} = {n_i^2}/{n_e}$ is used on electrode/semiconductor interfaces and the homogeneous Robin boundary condition $\hat{ \mathbf{n} } \cdot { \mathbf{J}_{e(h)}} = 0$ is used on semiconductor/insulator interfaces~\cite{Vasileska2010}. For Poisson equation, the Dirichlet boundary condition $\varphi  = {V_{el}} + {V_T} \ln{(n_e^s/n_i)}$ is enforced on electrode surfaces and the homogeneous Neumann boundary condition $\hat{\mathbf{n}} \cdot \nabla \varphi  = 0$ is used to truncate the computation domain~\cite{Vasileska2010}. Here, ${V_{el}}$ is the electric potential on the electrode, and ${V_{el}}=0$ for the negative electrode and ${V_{el}}=V_{\mathrm{bias}}$ for the positive one. For Maxwell equations, the computation domain is truncated with PMLs backed with PEC~\cite{Chen2020pml, Li2014DGTDBI}.

	The simulation domains are discretized using tetrahedrons (Fig.~\ref{PM3Dschem}). The minimum and the maximum edge lengths in the mesh are $10~\mathrm{nm}$ and $200~\mathrm{nm}$, respectively. The numbers of elements are ${1\,107\,866}$ and $13\,591$ in the full-device and unit-cell simulations, respectively. The tolerance of the Gummel iteration is $10^{-5}$ and the solution typically converges after $150$ iterations. The linear systems are solved using the GMRES iterative solver and an ILU preconditioner is reused throughout the Gummel iterations~\cite{Chen2020steadystate}. The physical duration of the transient stage is $8~\mathrm{ps}$ and the time-step sizes for the Maxwell and DD equations are $4\times{10^{-7}}~\mathrm{ps}$ and $2 \times {10^{-6}}~\mathrm{ps}$, respectively. {\color{black} These time-step sizes are chosen based on the Courant-Friedrichs-Lewy (CFL) conditions for the Maxwell and the DD equations~\cite{Chen2019multiphysics}, and are much smaller than the relaxation time ($10^{-3}~\mathrm{ps}$) of the carrier response on the PCD~\cite{Vasileska2010}.} Table~\ref{cost} provides the computational times (measured in ``core-hour'') required by the unit-cell and full-device simulations to complete the steady state and the transient stage. Note that the unit-cell simulation can be carried out on a workstation, while the full-device simulation requires $\sim 1~\mathrm{TB}$ of memory and calls for a parallelized solver and a distributed-memory system. The unit-cell scheme's steady state and transient stage simulations are $1800$ and $1500$ times faster.  
	
	{\color{black} The linear systems solved for the steady-state simulation are sparse however there is a trade-off between the ILU preconditioner sparsity and the number of GMRES iterations (denser preconditioner means smaller number of iterations). Therefore, the computational cost of the steady-state simulation is estimated to be between $O(N_{\mathrm{it}}N^2)$ and $O(N_{\mathrm{it}}N)$, where $N_{\mathrm{it}}$ is the number of GMRES iterations and $N$ is the number of unknowns. Note that $N_{\mathrm{it}}$ is larger for the full-device steady-state simulation since the mesh is more non-uniform (worsening the conditioning of the matrix systems). This short computational complexity analysis explains the large difference between the computation times required by the full-device and the unit-cell steady-state simulations.}
	
	{\color{black} The computation domain of the full-device transient-stage simulation contains many elements in the air background and in the PMLs. During time marching, only EM fields are updated for these elements while both EM fields and carrier densities are updated for the elements in the photoconductive region. Note that the exponentially decaying boundary layer of carrier densities require fine meshes on the boundaries of the photoconductive region. These fine meshes are required on almost all six faces of the photoconductive region for the full-device computation domain while in the unit-cell computation domain, fine meshes are only required on the top and bottom boundaries. These differences in the meshes used for full-device and unit-cell computation domains make the parallel load-balancing of the full-device transient-stage simulation significantly less efficient than that of the unit-cell transient-stage simulation. Therefore, even though the computational cost of the transient-stage simulation theoretically scales with $O(N)$ (linear in the number of unknowns $N$), the measured computation time comparison shows that the full-device transient-stage simulation is much slower than 80 times (the ratio of the numbers of unknowns in full-device and unit-cell computation domains).}
	
\begin{figure}[!b]
		\centering
		\subfloat[\label{PM3Dfulla}]{\includegraphics[width=0.8\columnwidth]{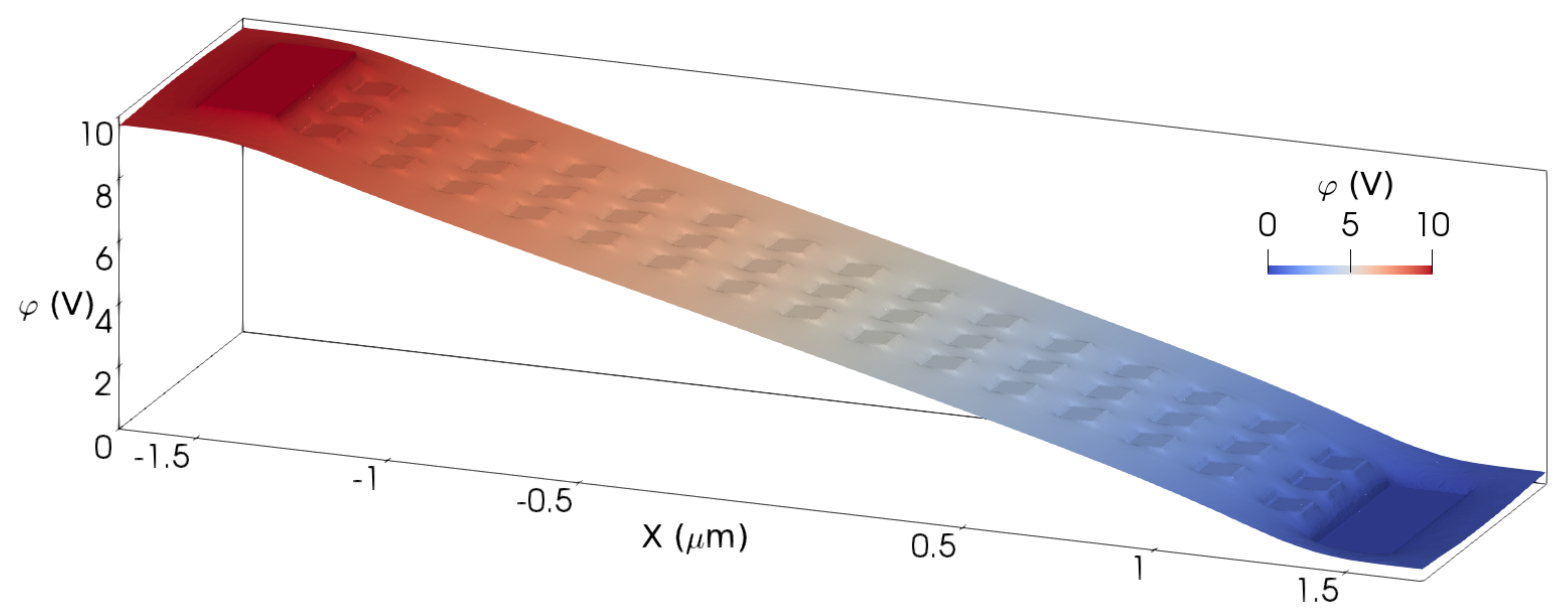}}\\
		\subfloat[\label{PM3Dfullb}]{\includegraphics[width=0.8\columnwidth]{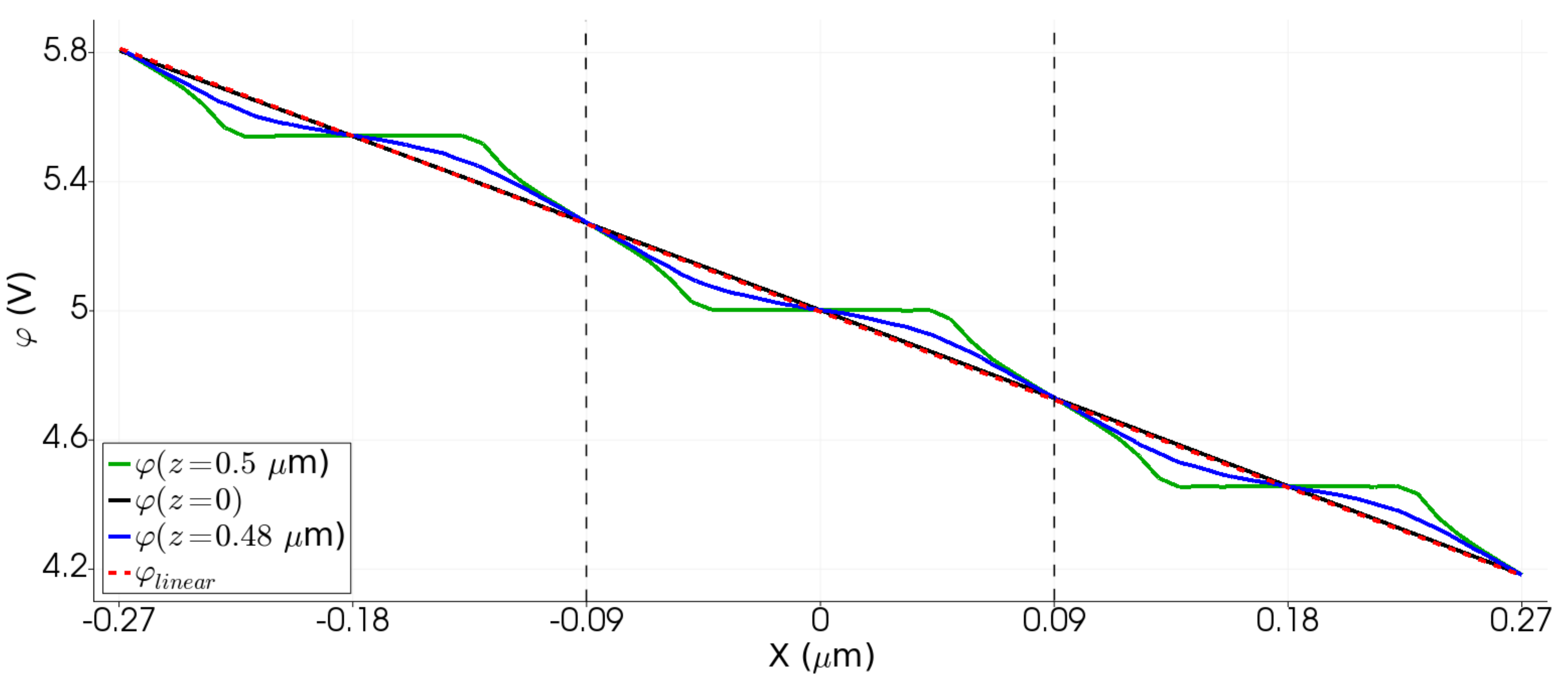}}\\
		\subfloat[\label{PM3Dfullc}]{\includegraphics[width=0.8\columnwidth]{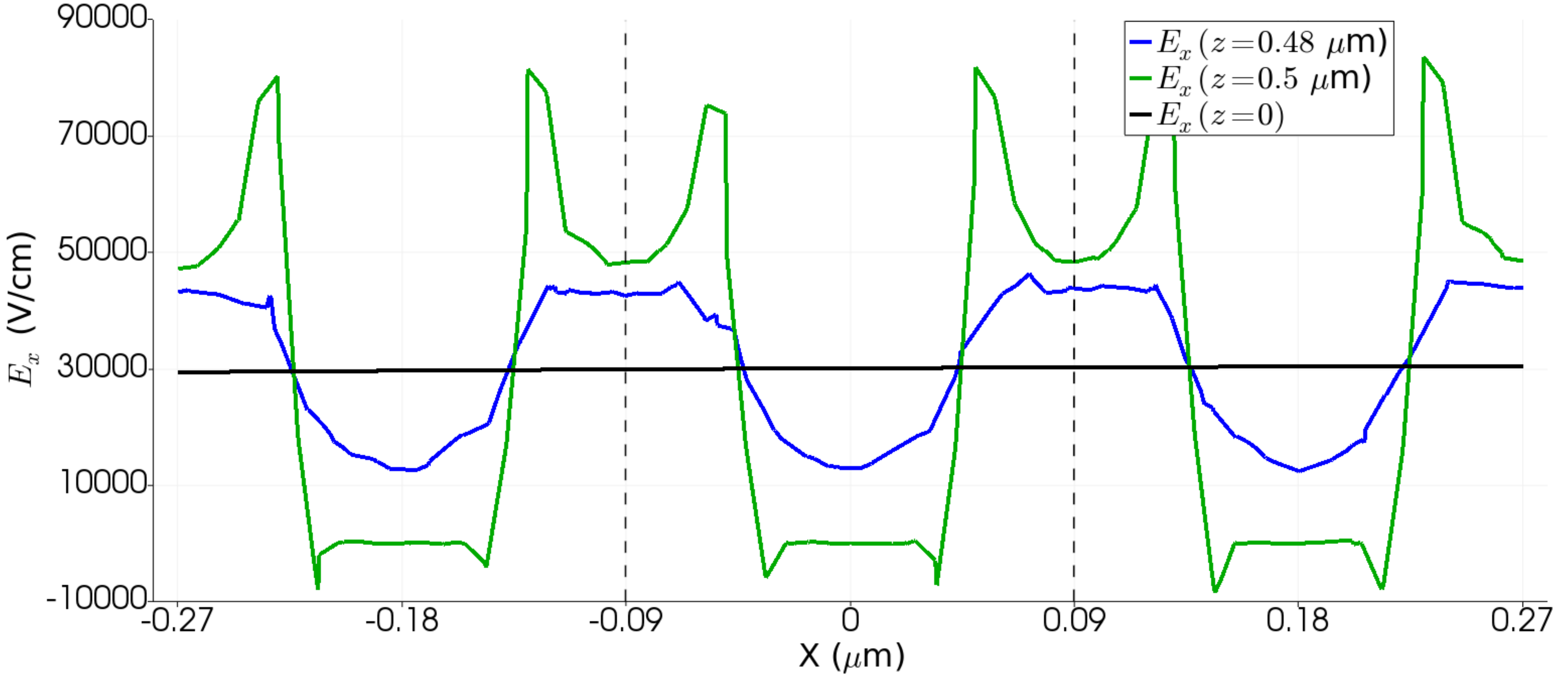}}
		\caption{ (a) Electric potential computed using the full-device simulation at the interface between the nanostructure and the photoconductive region ($z=0.5~\mu\mathrm{m}$). (b) Electric potential and (c) electric field on lines $(y=0,z=0.5~\mu\mathrm{m})$, $(y=0,z=0.48~\mu\mathrm{m})$ and $(x,y=0,z=0)$ near the device center. $\varphi_{linear}$ indicates the linear estimation of the potential drop.}
		\label{PM3Dfull}
	\end{figure}

	Fig.~\ref{PM3Dfull} shows the solutions obtained by the full-device simulation. Fig.~\ref{PM3Dfull} (a) shows the electric potential distribution on the interface between the nanostructures and the photoconductive region. One can clearly see that the potential drops equally across each unit cell and the local variations in all unit cells are approximately the same. Figs.~\ref{PM3Dfull} (b) and (c) show the electric potential and electric field on several lines along the $x$ direction, respectively. The dash lines mark the positions of the unit-cell surfaces. Fig.~\ref{PM3Dfull} (b) shows that, although the potential distributions at different $z$ positions are different, the potential drops are approximately the same. The linear drop estimation agrees with the solution very well on all unit-cell surfaces. Fig.~\ref{PM3Dfull} (c) shows the bias electric field is periodic as we expected.

\begin{figure}[!b]
		\centering
		\vspace{-0.2cm}
		\subfloat[\label{PM3Dunita}]{\includegraphics[height=0.3\columnwidth]{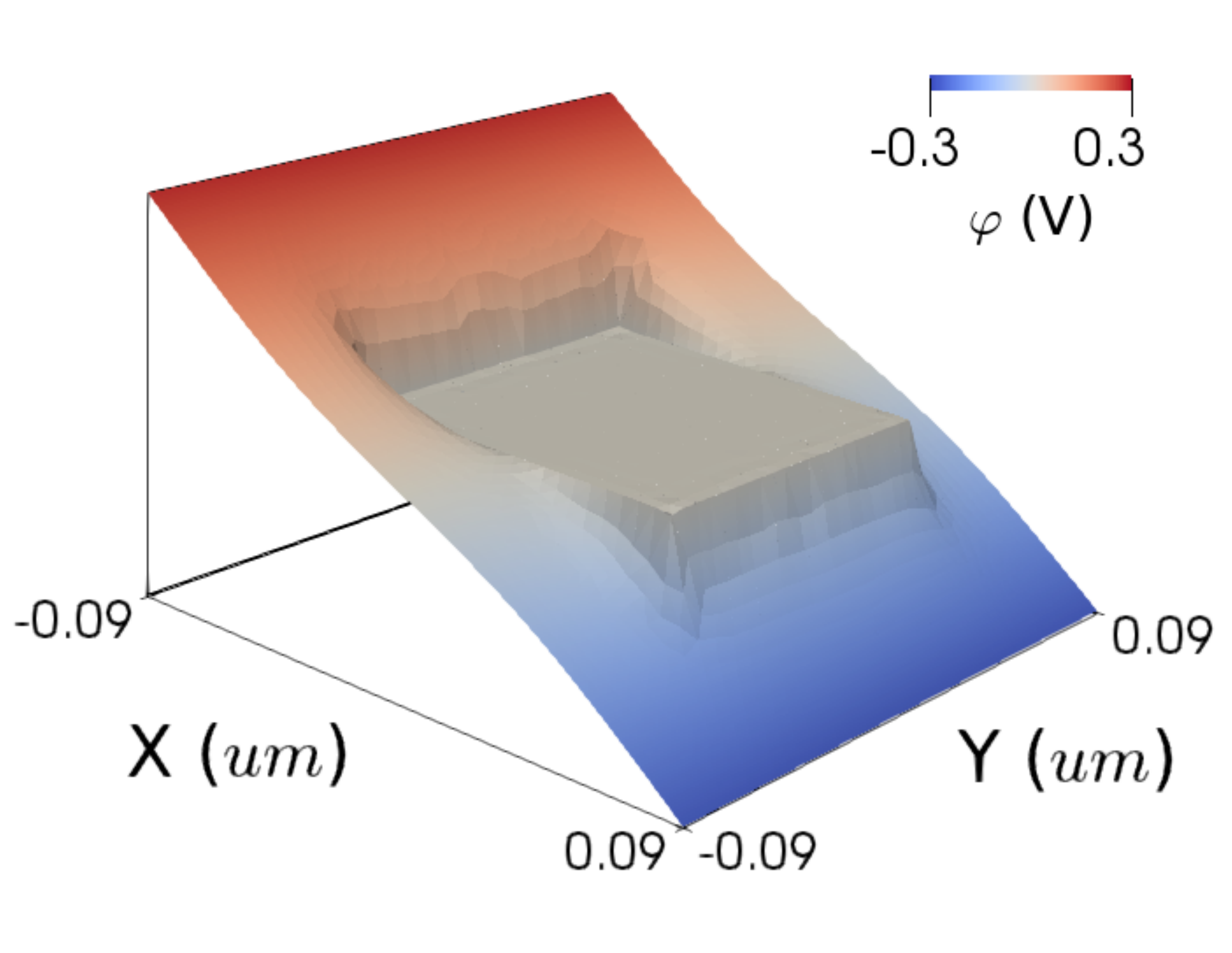}}\hspace{0.1cm}
		\subfloat[\label{PM3Dunitb}]{\includegraphics[height=0.33\columnwidth]{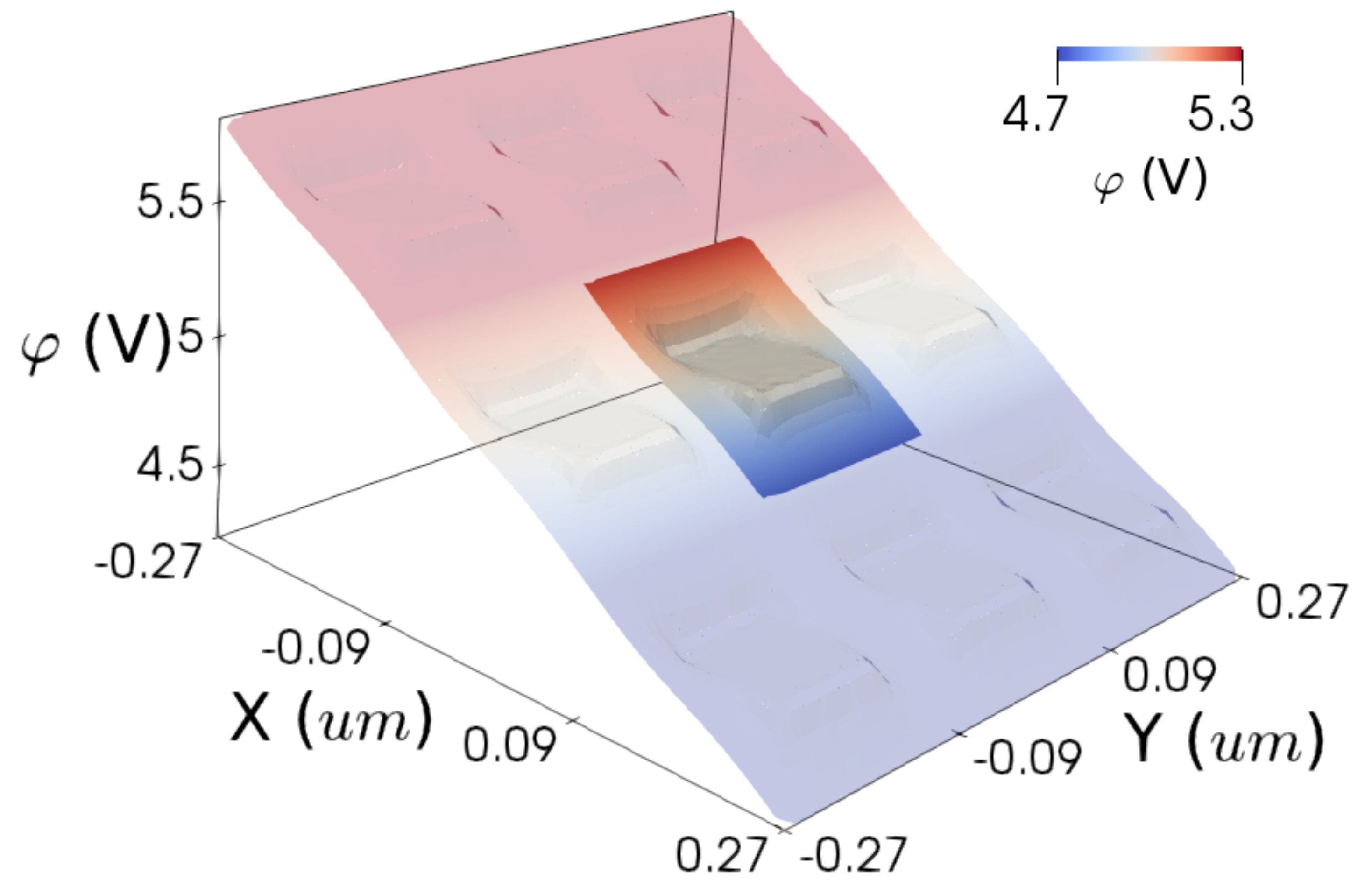}}\\
		\subfloat[\label{PM3Dunite}]{\includegraphics[height=0.45\columnwidth]{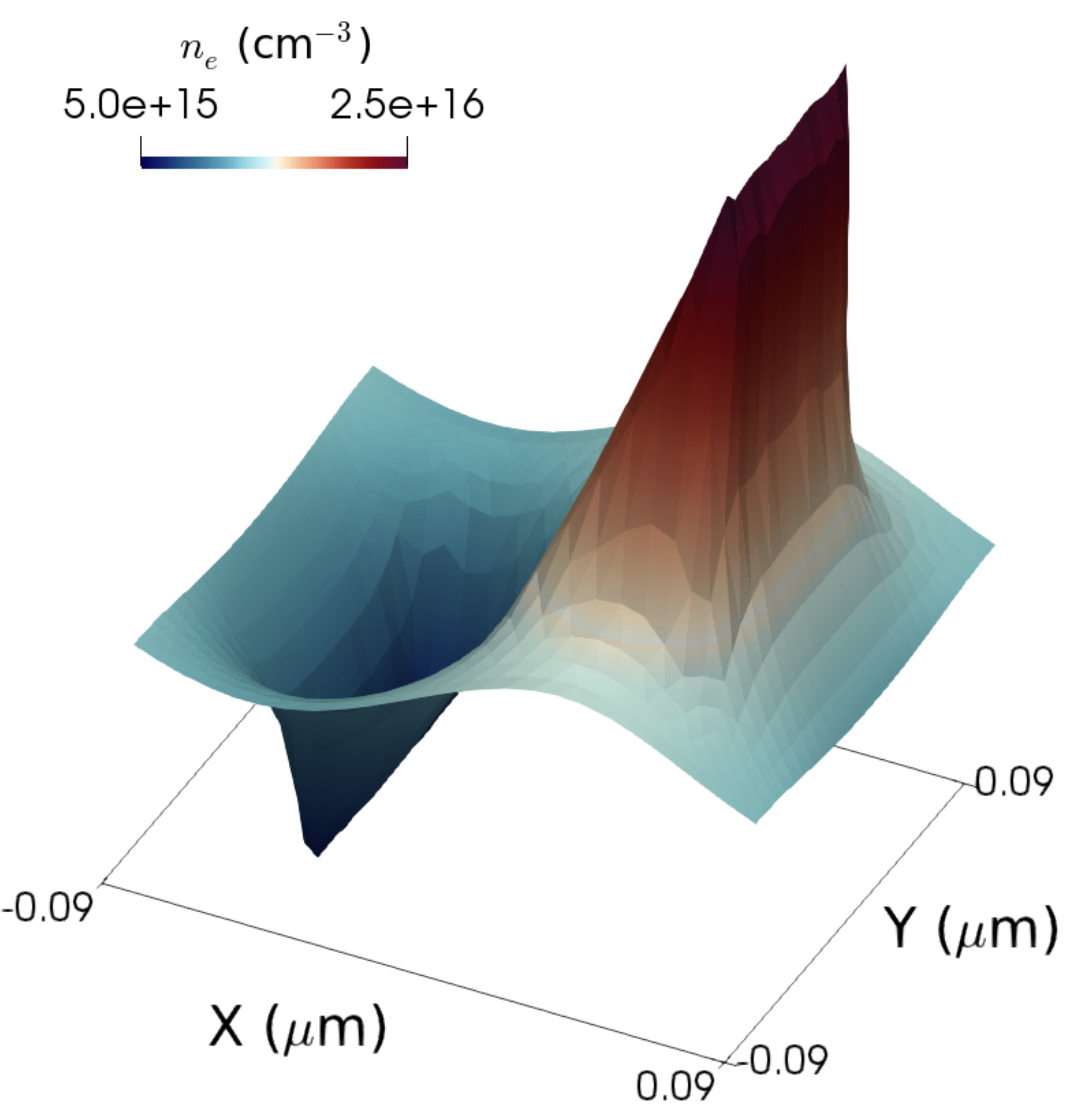}}\hspace{0.2cm}
		\subfloat[\label{PM3Dunitf}]{\includegraphics[height=0.4\columnwidth]{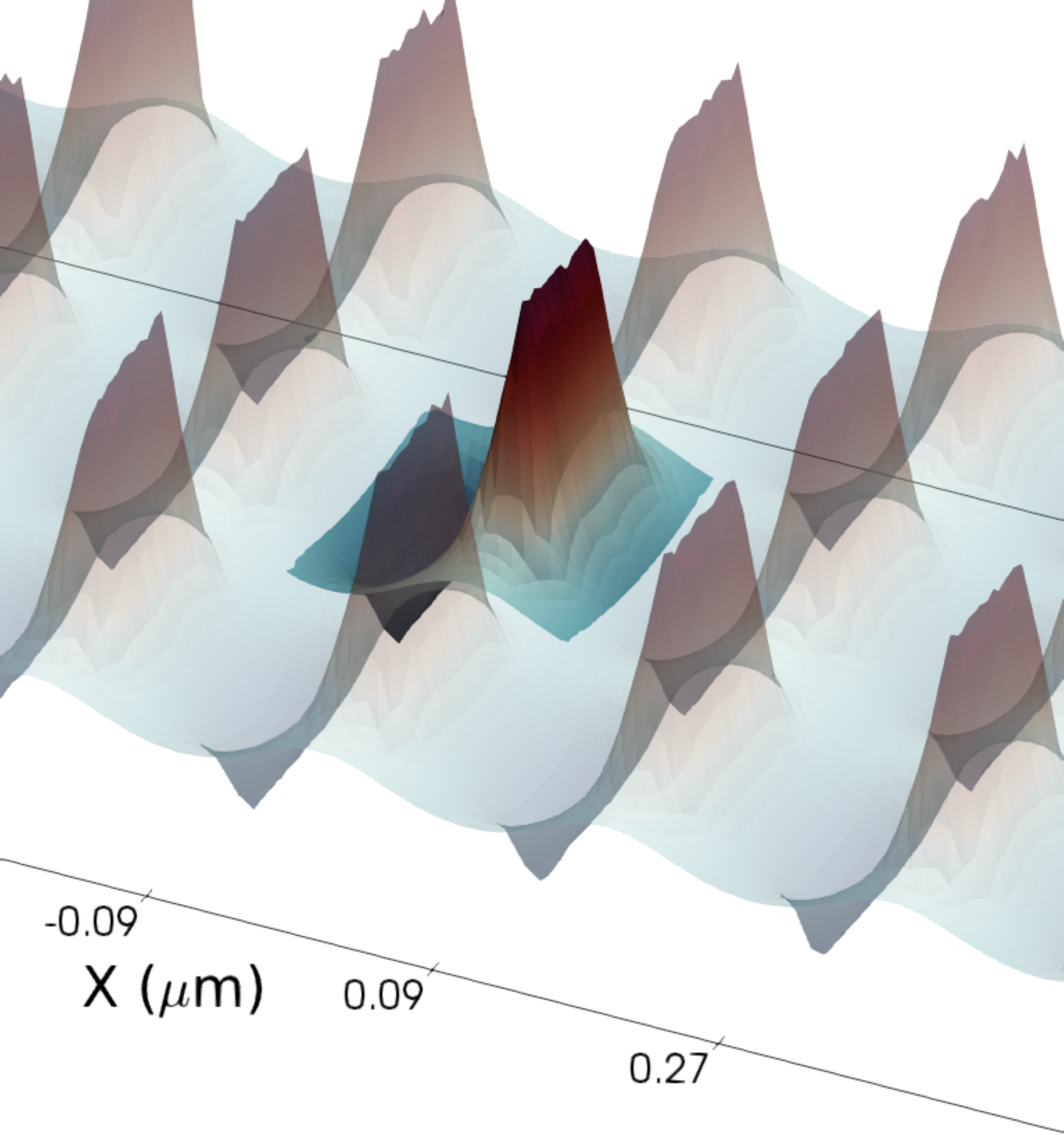}}\hspace{0.1cm}
		\caption{ (a) Electric potential and (c) electron density computed using the unit-cell scheme. (b) Electric potential and (d) electron density computed using the full-device simulation. The displayed solutions are extracted from the interface at $z=0.5~\mu\mathrm{m}$.}
		\vspace{-0.1cm}
		\label{PM3Dunit}
	\end{figure}

	Figs.~\ref{PM3Dunit} (a) and (c) show the steady-state electric potential and electron density calculated from the unit-cell scheme, respectively. For comparison, Figs.~\ref{PM3Dunit} (b) and (d) show those calculated using the full-device, where the solutions are set transparent (except those on the center unit cell) for better visualization and comparison. Very good agreement between two sets of results is observed. From Figs.~\ref{PM3Dunit} (a) and (b), one can see that although only the potential difference between boundaries is used in the unit-cell scheme, the potential variation inside the unit cell is same as that obtained from the full-device simulation. Since the bias electric fields are the same in all unit cells (the electric potential in different unit cells only differs by a constant), the mobility and the carrier densities are periodic. Fig.~\ref{PM3Dunit} (d) shows the electron density distribution in the full-device. And, Fig.~\ref{PM3Dunit} (c) shows that the solution obtained from the unit-cell scheme is same as that obtained from the full-device simulation. 

\begin{figure}[!t]
		\centerline{\includegraphics[width=0.9\columnwidth]{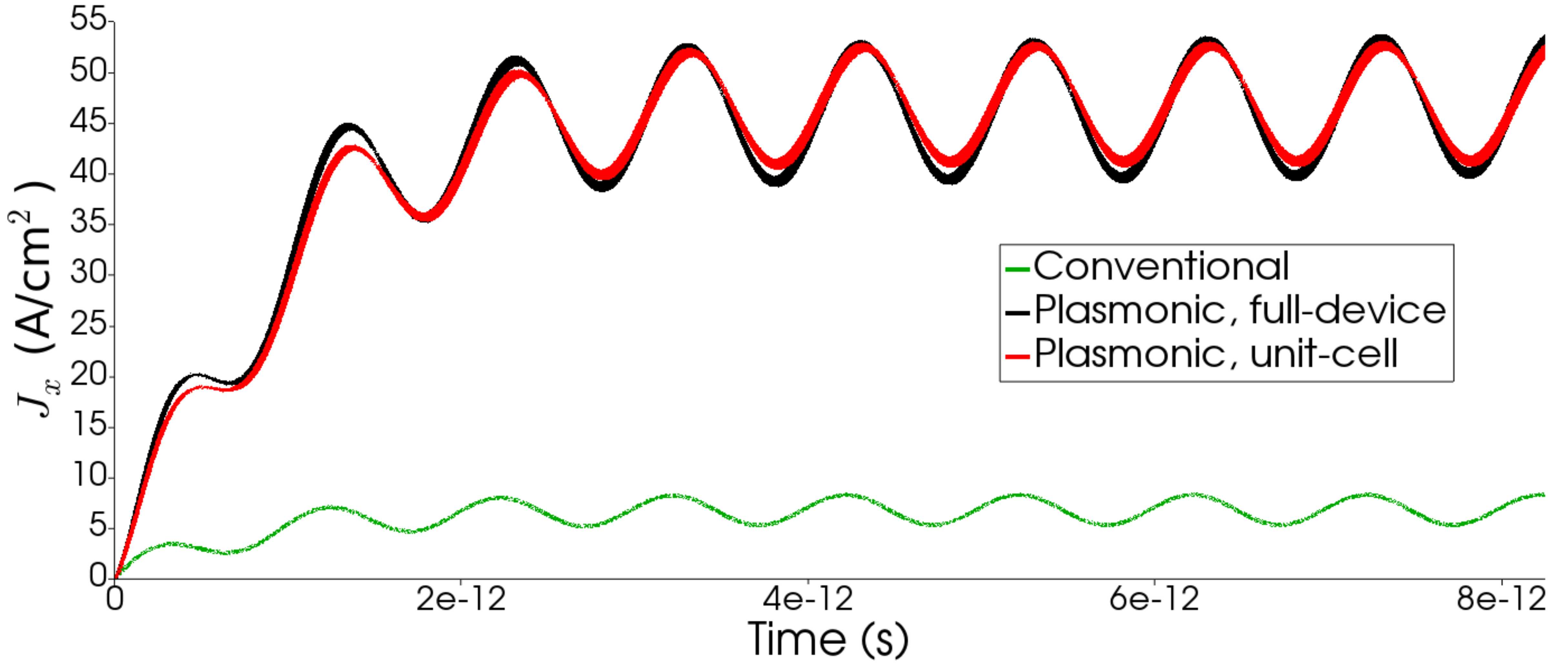}}
		\caption{The $x$ component of the transient current density obtained from the unit-cell and the full-device simulations.}
		\label{Current}
	\end{figure}

	Fig.~\ref{Current} compares the transient current densities obtained from the unit-cell and full-device simulations. The results agree well. To demonstrate the effect of the nanostructure on the device output, the current density obtained on the same device but without the nanostructure is also shown in the figure. It is clear that the photocurrent density increases significantly with the introduction of the plasmonic nanostructure yielding an enhancement factor of $5.9$.
	
	Note that the unit-cell scheme assumes an infinitely large optical pump aperture (the optical pump is same for all unit cells). In the full-device simulation, the pump beam and the device have finite widths, which results in two effects: (1) The pump power near the center cells is higher than that near the boundary cells and (2) the optical field is scattered by the electrodes and the $x$/$y$ boundaries of the device. These effects lead to a small difference between the transient current densities obtained by the two methods. 
	
	\section{Conclusion}
	\label{Conclusion}
	The large scale of 3D nanostructured PCDs and various multiscale/multiphysics and nonlinearly coupled physical phenomena involved in their operation render their direct simulation computationally very costly. The unit-cell scheme developed in this work dramatically reduces this computational cost, while the complex nonlinear EM/carrier interactions are still accurately accounted for. This scheme solves coupled systems of Poisson and stationary DD and time-dependent Maxwell and DD equations in a unit cell which represents one period of the nanostructure. The coupled equations systems are discretized using DG schemes. A PDBC is enforced on the unit-cell surfaces for Poisson equation, while PBCs are used for stationary and time-dependent DD equations and time-dependent Maxwell equations. These BCs are weakly enforced using the numerical flux of the DG methods. Numerical results demonstrate that the proposed unit-cell DG scheme maintains the accuracy of the DG scheme that operates on the full device while it is more than $1500$ times faster. 
	
	The unit-cell scheme developed in this work can be used in the design of PCDs not only with nanostructures but also antireflection layers and substrates. It can also be extended to account for organic devices operating with possibly more than two types of carriers.

	\section*{Acknowledgment}
	This research is supported in part by the King Abdullah University of Science and Technology (KAUST) Office of Sponsored Research (OSR) under Award No 2016-CRG5-2953 and in part by the Okawa Foundation Research Grant. The authors would like to thank the KAUST Supercomputing Laboratory (KSL) for providing the required computational resources.


	

	
\end{document}